\documentclass[11pt]{article}

\renewcommand{\baselinestretch}{1.1}

\usepackage{amsmath,amsthm, amssymb}

\newcommand{\dst}{\displaystyle}
\newcommand{\ba}{\begin{array}}
\newcommand{\ea}{\end{array}}

\newcommand{\bea}{\begin{eqnarray}}
\newcommand{\eea}{\end{eqnarray}}
\newcommand{\bma}{\begin{matrix}}
\newcommand{\ema}{\end{matrix}}
\newcommand{\bpm}{\begin{pmatrix}}
\newcommand{\epm}{\end{pmatrix}}
\newcommand{\nn}{\nonumber}
\newcommand{\half}{\frac{1}{2}}
\newcommand{\qter}{\frac{1}{4}}
\newcommand{\mc}{\mathcal}

\newcommand{\p}{\partial}

\newcommand{\rr}{\prime}
\newcommand{\ov}{\overline}

\newcommand{\wt}{\widetilde}
\newcommand{\Psibar}{{\ov \Psi}{}}

\newcommand{\tabar}{\ov \ta}
\newcommand{\phibar}{\ov\phi{}}

\newcommand{\eps}{\varepsilon}
\newcommand{\ep}{\epsilon}
\newcommand{\al}{\alpha}
\newcommand{\la}{\lambda}
\newcommand{\da}{\delta}
\newcommand{\om}{\omega}
\newcommand{\Ga}{\Gamma}
\newcommand{\ga}{\gamma}
\newcommand{\La}{\Lambda}
\newcommand{\si}{\sigma}
\newcommand{\ta}{\theta}

\newcommand{\epbar}{{\ov\ep}}

\newcommand{\Abar}{{\ov A}{}}
\newcommand{\betabar}{{\ov\beta}}

\newcommand{\Om}{\Omega}

\newcommand{\un}{\underline}
\newcommand{\mfa}{\mathsf{a}}
\newcommand{\mfb}{\mathsf{b}}
\newcommand{\mfc}{\mathsf{c}}

\title{\bf Mass-Deformed BLG Theory\\ in Light-Cone Superspace}
\author{\\[1cm]
 \large{\bf Dmitry V.~Belyaev}~\thanks{belyaev@phys.ufl.edu}\\[1cm]
 \it Institute for Fundamental Theory,\\ 
 \it Department of Physics, University of Florida\\
 \it Gainesville, FL 32611, USA}
\date{}

\begin{document}

\numberwithin{equation}{section}

\maketitle
\thispagestyle{empty}

\begin{abstract}
Maximally supersymmetric mass deformation of the Bagger-Lambert-Gustavsson (BLG) theory corresponds to a \emph{non-central} extension of the $d=3$ $N=8$ Poincar\'e superalgebra (allowed in three dimensions). We obtain its light-cone superspace formulation which has a novel feature of the dynamical supersymmetry generators being \emph{cubic} in the kinematical ones. The mass deformation picks a quaternionic direction, described by $\Om_m{}^n$, which breaks the $SO(8)$ $R$-symmetry down to $SO(4)\times SO(4)$. The Hamiltonian of the theory is shown to be a quadratic form of the dynamical supersymmetry transformations, to all orders in the mass parameter, $M$, and the structure constants, $f^{a b c d}$.
\end{abstract}

\newpage

\noindent\rule\textwidth{.1pt}		
\tableofcontents
\vspace{20pt}
\noindent\rule\textwidth{.1pt}

\setcounter{page}{1}

\section{Introduction}

The Bagger-Lambert-Gustavsson (BLG) theory \cite{bl1,gus,bl2} is the maximally supersymmetric and superconformal three-dimensional gauge theory of Chern-Simons type. It has rigid symmetry described by the superconformal group $OSp(2,2|8)$ and local symmetry described by the 3-Lie algebra with structure constants $f^{b c d}{}_a$. This theory has a very interesting mass deformation \cite{gsp,hll}, which breaks the conformal symmetries while preserving all of the maximal supersymmetry. The resulting symmetry group is a \emph{non-central} extension of the $d=3$ $N=8$ superPoincar\'e group \cite{nahm,lm}. Such an extension is forbidden in four and higher dimensions \cite{cm,hls}, which makes this three-dimensional theory very special.

As is well-known, supersymmetric theories enjoy living in superspace. The latter comes in two varieties: off-shell superspace \cite{ss} (best suited for minimal supersymmetry) and on-shell superspace \cite{sg} (best suited for maximal supersymmetry \cite{bln1}). The former requires additional auxiliary fields, whereas the latter operates with only physical degrees of freedom. Light-cone (LC) superspace is the best known example of on-shell superspace. It has been used, in particular, to prove the UV finiteness of the $N=4$ super-Yang-Mills \cite{man,bln2}. One surprising feature of the LC superspace is that the LC superspace Hamiltonian of a maximally supersymmetric theory (in all the cases studied to date) appears to be a quadratic form of the dynamical supersymmetry transformation of the basic superfield \cite{abkr,abhs,db,bbkr}. In the LC superspace formulation of the BLG theory \cite{bn,db,bbkr}, this has been verified \cite{bbkr} to linear order in $f^{a b c d}$. In this paper, we will prove that this property holds to all orders in $f^{a b c d}$.

Attempting to construct the LC superspace formulation of the mass-deformed BLG theory from scratch, by solving the constraints imposed by the symmetry group (as in \cite{bbkr}), one would encounter (at least) two problems. First, the dynamical supersymmetry generators, $\mc{Q}$'s, must be \emph{cubic} in the kinematical supersymmetry generators, $q$'s, because in the commutator of two supersymmetries one must find an $R$-symmetry transformation which is quadratic in $q$'s. At the same time, in the commutator of two $\mc{Q}$'s there should be no terms quartic in $q$'s, as no such symmetry generators exist. The apparent quartic terms must somehow cancel. Second, the mass deformation should break the $SO(8)$ $R$-symmetry of the BLG theory down to $SO(4)\times SO(4)$. And it is not a priori obvious how to accomplish this in the LC superspace setting best suited for keeping $SU(4)$ $R$-symmetry manifest.

In this paper, we will perform the top-down reduction \cite{db} of the known (covariantly formulated) mass-deformed BLG theory \cite{gsp,hll} to its LC superspace form. This allows us to solve the above mentioned problems. In the covariant formulation, the mass parameter is accompanied by the 32$\times$32 matrix $\Ga_{3456}$ (the product of four 11-dimensional gamma matrices) which in our conventions is
\bea
\Ga_{3456}=i\bpm {\bma + & \\ & - \ema} & \\ & {\bma + & \\ & - \ema} \epm
\otimes
\bpm \Om^m{}_n & \\ 0 & \Om_m{}^n \epm \ ,
\eea
where $\Om^m{}_n=-\Om_n{}^m$ is one of the quaternionic matrices in the algebra of $SO(4)\subset SU(4)$. We will see that this matrix intertwines the $SO(8)$ $R$-symmetry generators \cite{bbkr} ($T_m^n$, $T_{m n}$, $T^{m n}$ and $T$) in a way that reduces the $R$-symmetry group to $SO(4)\times SO(4)$. We will also see that $\mc{Q}$'s are indeed cubic in $q$'s, whereas the following identity
\bea
\da_{[n}^m\Om_k{}^l\Om_r{}^s q_t q_{s]}=0
\eea
is responsible for the absence of terms quartic in $q$'s inside the commutator of two $\mc{Q}$'s.

We will see that the mass deformation affects the dynamical supersymmetry transformations, but not the kinematical ones. (Therefore, mass is treated as an interaction in LC superspace.) The modification is fairly simple, given that the mass parameter, $M$, never multiplies the structure constants, $f^{b c d}{}_a$. The modification at the Lagrangian level is more involved, but once we show that the LC superspace Hamiltonian of the mass-deformed BLG theory is a quadratic form of the dynamical supersymmetry transformations, the overall structure becomes quite simple.

In Section \ref{sec-abelianBLG}, we analyze the mass-deformed BLG theory with $f^{b c d}{}_a$ set to zero (i.e. the ``abelian'' version of the theory) and with the gauge indices on the fields accordingly suppressed. The indices will be reintroduced together with $f^{b c d}{}_a$ in Section \ref{sec-fullBLG}. Some technical details (including the major part of the proof of the quadratic form property of the LC superspace Hamiltonian) are delegated to the Appendices. Throughout the paper, we follow the conventions of \cite{db}.

\section{The abelian theory}
\label{sec-abelianBLG}

In this section, we will analyze the mass-deformed BLG theory \cite{gsp,hll} with $f^{b c d}{}_a=0$. In the covariant formulation, the field content of the theory consists of eight scalars, $X^I$, $I=3,\dots,10$, and a single 32-component Majorana spinor, $\Psi$, satisfying an additional constraint $\Ga_{012}\Psi=-\Psi$. On-shell, there are 8 bosonic and 8 fermionic degrees of freedom. They fit nicely into the LC superfield $\phi$ \cite{bln1}. Our goal is to find how $\phi$ varies under supersymmetry transformations. In the covariant formulation, the latter are described by a 32-component Majorana spinor, $\ep$, satisfying $\Ga_{012}\ep=+\ep$. In the LC superspace formulation, $\ep$ is reduced to 8 kinematical supersymmetry parameters, $\al^m$ and $\al_m=(\al^m)^\ast$, and 8 dynamical supersymmetry parameters, $\beta^m$ and $\beta_m=(\beta^m)^\ast$; $m=1,2,3,4$. We will see that the mass deformation affects only the latter, and that the mass parameter, $M$, appears multiplied by a matrix $\Om_m{}^n$ that breaks the $SO(8)$ $R$-symmetry down to $SO(4)\times SO(4)$. We will close this section with the discussion of the Hamiltonian, $H$, and the dynamical Lorentz boost generator, $\mc{J}^{-}$.

\subsection{Covariant formulation}

The action of the abelian mass-deformed BLG theory is $S=\int d^3x\mc{L}$ with
\bea
\label{LwithM}
\mc{L}=-\half(\p_\mu X^I)(\p^\mu X^I)+\frac{i}{2}\Psibar\Ga^\mu\p_\mu\Psi
-\half M^2 X^I X^I+\frac{i}{2}M\Psibar\Ga_{3456}\Psi \ ,
\eea
where $\mu=0,1,2$ and $I=3,4,5,6,7,8,9,10$. The symmetries of this action are
\begin{itemize}
\item
translations, with parameters $v^\mu$,
\bea
\da_v X^I=v^\mu\p_\mu X^I, \quad
\da_v\Psi=v^\mu\p_\mu\Psi \ ;
\eea
\item
Lorentz transformations, with parameters $\la^{\mu\nu}=-\la^{\nu\mu}$,
\bea
\label{Lorentz}
\da_\la X^I = \la^{\mu\nu}x_\mu\p_\nu X^I, \quad
\da_\la\Psi = \la^{\mu\nu}x_\mu\p_\nu\Psi+\qter\la^{\mu\nu}\Ga_{\mu\nu}\Psi \ ;
\eea
\item
$SO(4)\times SO(4)$ $R$-symmetry transformations, 
\bea
\da_R X^i=R^{i j}X^j, \quad
\da_R X^{i^\rr}=R^{i^\rr j^\rr}X^{j^\rr}, \quad
\da_R\Psi=\qter R^{i j}\Ga_{i j}\Psi+\qter R^{i^\rr j^\rr}\Ga_{i^\rr j^\rr}\Psi \ ,
\eea
where $i=3,4,5,6$ and $i^\rr=7,8,9,10$; the parameters $R^{i j}=-R^{j i}$ parametrize the first $SO(4)$, and $R^{i^\rr j^\rr}=-R^{j^\rr i^\rr}$ parametrize the second $SO(4)$;
\item
supersymmetry transformations, with the parameter $\ep$,
\bea
\label{susy}
\da_\ep X^I = i\epbar\Ga^I\Psi, \quad
\da_\ep \Psi = \Ga^\mu\Ga^I\ep\p_\mu X^I-M\Ga_{3456}\Ga^I\ep X^I \ .
\eea
\end{itemize}
The algebra of these symmetries closes on-shell, i.e. provided the equations of motion implied by (\ref{LwithM}),
\bea
\label{eom}
(\p^\mu\p_\mu-M^2)X^I=0, \quad
(\Ga^\mu\p_\mu+M\Ga_{3456})\Psi=0 \ ,
\eea
are satisfied. The key commutator is that of two supersymmetries, for which we find
\bea
[\da_{\ep_1},\da_{\ep_2}]=\da_v+\da_R \ ,
\eea
where (note that our conventions are such that $\Ga_{789(10)}=-\Ga_{3456}\Ga_{012}$)
\bea
v^\mu=-2i(\epbar_2\Ga^\mu\ep_1), \quad
R^{i j}=2i M(\epbar_2\Ga^{i j}\Ga_{3456}\ep_1), \quad
R^{i^\rr j^\rr}=2i M(\epbar_2\Ga^{i^\rr j^\rr}\Ga_{789(10)}\ep_1) \ .
\eea
As the commutator of two supersymmetry transformations yields the $R$-symmetry transformation (in addition to the standard translation), this algebra is a \emph{non-central} extension of the $d=3$ $N=8$ superPoincar\'e algebra.~\footnote{
This superalgebra (with 16 supercharges) is described in Appendix E.2 of \cite{lm}. It is a doubled version of the superalgebra with 8 supercharges given explicitly in Appendix E.1 of \cite{lm} and described after eq.~(48) in \cite{nahm}.
}

\subsection{LC supersymmetry transformations: component form}

Using the LC projectors, $P_{+}=-\half\Ga_{+}\Ga_{-}$ and $P_{-}=-\half\Ga_{-}\Ga_{+}$, we define $\ep_{\pm}=P_{\pm}\ep$ and $\Psi_{\pm}=P_{\pm}\Psi$ \cite{db}. The fermionic equation of motion in (\ref{eom}) can then be used to solve for $\Psi_{-}$,
\bea
\label{PsiMinus}
\Psi_{-}=\frac{1}{2\p^{+}}\Ga_{-}(\Ga_2\p+M\Ga_{3456})\Psi_{+} \ ,
\eea
where $\p^{+}=\frac{1}{\sqrt2}(-\p_0+\p_1)$ and $\p=\p_2$.
From now on, only variations of $\Psi_{+}$ need to be considered. The transformations (\ref{susy}) split into the kinematical supersymmetry transformations,
\bea
\label{ksusy}
\da_{\ep_{-}}X^I &=& i\epbar_{-}\Ga^I\Psi_{+} \nn\\
\da_{\ep_{-}}\Psi_{+} &=& \Ga_{+}\Ga^I\ep_{-}\p^{+}X^I \ ,
\eea
and the dynamical supersymmetry transformations,
\bea
\label{dsusy}
\da_{\ep_{+}}X^I &=& \frac{i}{2\p^{+}}\epbar_{+}\Ga^I\Ga_{-}(\Ga_2\p+M\Ga_{3456})\Psi_{+} \nn\\
\da_{\ep_{+}}\Psi_{+} &=& (\Ga_2\p-M\Ga_{3456})\Ga^I\ep_{+}X^I \ .
\eea
As the mass deformation does not affect the kinematical supersymmetry transformations (\ref{ksusy}), the fitting of the degrees of freedom into the LC superfield $\phi$ is as in \cite{db}. The eight scalars $X^I$ define the bosonic components of $\phi$ as follows,
\bea
\label{AandC}
A &=& \frac{1}{\sqrt2}(X^3+i X^4) \nn\\
C_{m n} &=& \frac{1}{\sqrt2}(\eta_1 X^5+\eta_2 X^6+\eta_3 X^7)
+\frac{i}{\sqrt2}(\wt\eta_1 X^8+\wt\eta_2 X^9+\wt\eta_3 X^{10}) \ ,
\eea
where $(\eta_\mfa)_{m n}$ and $(\wt\eta_\mfa)_{m n}$, $\mfa=1,2,3$, are six 4$\times$4 matrices (the so-called 't Hooft symbols \cite{thooft,bvvn}) whose explicit form and properties are given in Appendix~\ref{appTH}. The 8 fermionic components in $\Psi_{+}$ define the fermionic components of $\phi$ as follows,
\bea
\Psi_{+}=\bpm \psi^m_{+} \\ \psi_{m+} \epm, \quad
\psi^m_{+}=\bpm 0\\0\\\chi^m\\0 \epm, \quad
\psi_{m+}=\bpm 0\\\chi_m\\0\\0 \epm ; \quad
\chi_m=(\chi^m)^\ast \ .
\eea
The 8 fermionic parameters in $\ep_{-}$ define the kinematical supersymmetry parameters $\al$,
\bea
\ep_{-}=\bpm \ep^m_{-} \\ \ep_{m-} \epm, \quad
\ep^m_{-}=\bpm 0\\0\\0\\\al^m \epm, \quad
\ep_{m-}=\bpm -\al_m\\0\\0\\0 \epm ; \quad
\al_m=(\al^m)^\ast \ ,
\eea
whereas the 8 parameters in $\ep_{+}$ define the dynamical supersymmetry parameters $\beta$,~\footnote{
In \cite{db}, the parameters of the dynamical supersymmetry were called $\eta^m$. Now we use $\beta^m$ instead, in order to avoid confusion with the 't Hooft symbols, which are used extensively in this paper.
}
\bea
\ep_{+}=\bpm \ep^m_{+} \\ \ep_{m+} \epm, \quad
\ep^m_{+}=\bpm 0\\\beta^m\\0\\0 \epm, \quad
\ep_{m+}=\bpm 0\\0\\\beta_m\\0 \epm ; \quad
\beta_m=(\beta^m)^\ast \ .
\eea
The component form of the kinematical supersymmetry transformations is as in \cite{db},
\bea
\label{LCksusy}
\da_{\ep_{-}}A &=& i\sqrt2\al^m\chi_m \nn\\
\da_{\ep_{-}}\chi_m &=& 2\p^{+}\Big(A\al_m+C_{m n}\al^n\Big) \nn\\
\da_{\ep_{-}}C_{m n} &=& -i\sqrt2\Big(\al_m\chi_n-\al_n\chi_m+\eps_{m n k l}\al^k\chi^l\Big) \ ,
\eea
whereas in the dynamical supersymmetry transformations we find $O(M)$ modifications,
\bea
\label{LCdsusy}
\da_{\ep_{+}}A &=& -\frac{1}{\p^{+}}\Big(\beta^m\p\chi_m+i M\beta^m\Om_m{}^n\chi_n\Big) \nn\\
\da_{\ep_{+}}\chi_m &=& -i\sqrt2\p\Big(A\beta_m-C_{m n}\beta^n\Big)
-\sqrt2 M\Om_m{}^n\Big(A\beta_n-C_{n k}\beta^k\Big) \nn\\
\da_{\ep_{+}}C_{m n} &=& \frac{\p}{\p^{+}}\Big(-\beta_m\chi_n+\beta_n\chi_m+\eps_{m n k l}\beta^k\chi^l\Big) \nn\\
&-&\frac{i M}{\p^{+}}\Big((\beta_m\Om_n{}^k-\beta_n\Om_m{}^k)\chi_k-\eps_{m n k l}\beta^k\Om_s{}^l\chi^s\Big) \ ,
\eea
where we defined $\Om_m{}^n\equiv(\eta_3)_{m n}$ (see Appendix~\ref{appTH}), so that~\footnote{
In the matrix-to-second-rank-tensor correspondence, we assign the left index to the rows and the right index to the columns. Then, for example, $\Om_1{}^2=+1$ and $\Om_2{}^1=-1$. Also, $\Om_m{}^n=\Om^m{}_n$ expresses the reality of $\Om$, whereas $\Om_m{}^n=-\Om_n{}^m$ expresses its antisymmetry. For a symmetric matrix (such as $I_4$ corresponding to $\da_m{}^n$) there is no need to distinguish between left and right indices, and so we write simply $\da_m^n$.
}
\bea
\Om_m{}^n=\bpm {\bma & + \\ - & \ema} & \\ & {\bma & + \\ - & \ema} \epm
=(\Om_m{}^n)^\ast \equiv \Om^m{}_n=-\Om_n{}^m, \quad
\Om_m{}^k\Om_k{}^n=-\da_m^n \ .
\eea
Note also that $\Om_m{}^m=0$, which is frequently used in what follows.

\subsection{LC supersymmetry transformations: superfield form}

The component fields enter the superfield $\phi$ in such a way that \cite{db}
\bea
\label{phicomp}
&&\phi_|=\frac{1}{\p^{+}}A, \quad
d_m\phi_|=\frac{i}{\p^{+}}\chi_m, \quad
d_{m n}\phi_|=-i\sqrt2 C_{m n} \nn\\
&&d_{m n k}\phi_|=-\sqrt2\eps_{m n k l}\chi^l, \quad
d_{m n k l}\phi_|=2\eps_{m n k l}\p^{+}\Abar \ .
\eea
The superfield $\phi$ is chiral, $d^m\phi=0$, and satisfies the reality condition (``inside-out constraint'')
\bea
\label{inout}
\phibar\equiv\phi^\ast=\frac{d_{[4]}}{2\p^{+2}}\phi, \quad d_{[4]}\equiv d_1 d_2 d_3 d_4 \ .
\eea
The same constraints must then be satisfied by the supersymmetry variation superfield $\da_\ep\phi$. For the kinematical supersymmetry transformations, the answer is
\bea
\da_{\ep_{-}}\phi=\sqrt2(\al^m q_m-\al_m q^m)\phi \ .
\eea
It is easy to verify that this reproduces (\ref{LCksusy}) using that
\bea
\{q^m,q_n\}=Z\da^m_n, \quad{} 
\{d^m,d_n\}=-Z\da^m_n, \quad Z\equiv i\sqrt2\p^{+} \ ,
\eea
and that the $\ta=0$ projection (denoted by ``$|$'') of $q$'s is equal to the projection of $d$'s. 
For the dynamical supersymmetry, we write
\bea
\label{phidsusy}
\da_{\ep_{+}}\phi = (\eta^m Q_m-\eta_m Q^m)\phi \ ,
\eea
where $Q$'s must have a part linear in $q$'s and a part cubic in $q$'s. The superfield transformation should reproduce (\ref{LCdsusy}) upon projection. We already know the answer for the $M$-independent part of $Q$'s \cite{db}. For the $M$-dependent part, we choose the most general ansatz and then fix the coefficients accordingly. The final result of this analysis is
\bea
\label{QQbar}
Q_m &=& +i\frac{\p}{\p^{+}}q_m-\frac{M}{Z\p^{+}}(Z\Om_m{}^n q_n+\Om_m{}^n q^k q_{n k}-\Om_k{}^n q^k q_{n m}) \nn\\
Q^m &=& -i\frac{\p}{\p^{+}}q^m+\frac{M}{Z\p^{+}}(Z\Om_n{}^m q^n+\Om_n{}^m q_k q^{n k}-\Om_n{}^k q_k q^{n m}) \ .
\eea
where $q_{n k}\equiv q_n q_k$. Direct evaluation then yields
\bea
\label{QQPM}
\{Q^m,Q_n\}=i\sqrt2\frac{1}{\p^{+}}(\p^2-M^2)\da^m_n \ ,
\eea
which confirms the correctness of the result. The terms quartic in $q$'s cancel thanks to the following identity
\bea
\Big\{
\da^m_n(\Om_k{}^l\Om_t{}^s q_{l s}-q_{k t})
+2\Om_k{}^m(\Om_t{}^s q_{s n}-\Om_n{}^s q_{s t}) \qquad && \nn\\
+2\Om_n{}^m\Om_k{}^s q_{s t} 
+2\da_t^m(\Om_n{}^l\Om_k{}^s q_{l s}-q_{n k})
\Big\} q^{k t} &=& 0 \ ,
\eea
which is the expanded version of the obvious identity
\bea
\da^m_{[n}\Om_k{}^l\Om_t{}^s q_{l s]}q^{k t}=0 \ .
\eea

\subsection{$SO(4)\times SO(4)$ $R$-symmetry}

When $M=0$, the (anti)commutator of the dynamical and kinematical supersymmetry generators yields only the translation in the transverse direction. When $M\neq0$, we also find terms quadratic in $q$'s, which can be organized in terms of the $SO(8)$ $R$-symmetry group generators (of the $M=0$ theory) \cite{bbkr}
\bea
&&T^m_n=-\frac{1}{Z}(q^m q_n-\qter\da^m_n q^k q_k), \quad
T=-\frac{1}{4Z}(q^k q_k-q_k q^k) \nn\\
&&T^{m n}=-\frac{1}{Z}q^m q^n, \quad
T_{m n}=-\frac{1}{Z}q_m q_n \ .
\eea
Explicitly, we find that
\bea
\{q^m,Q_n\} &=& -\sqrt2\da^m_n\p+i\sqrt2 M S_n^m \nn\\
\{q_m,Q_n\} &=& i\sqrt2 M S_{m n}, \quad
\{q^m,Q^n\}=-i\sqrt2 M S^{m n} \ ,
\eea
where we defined
\bea
S_n^m &=& T_k^m\Om_n{}^k+T_n^k\Om_k{}^m-\da_n^m T_l^k\Om_k{}^l-T\Om_n{}^m \nn\\
S_{m n} &=& \Om_m{}^k T_{n k}-\Om_n{}^k T_{m k}, \quad
S^{m n}=\Om_k^m T^{n k}-\Om_k^n T^{m k} \ .
\eea
We will now show that these generators generate the $SO(4)\times SO(4)$ subgroup of $SO(8)$. We will see that $S_m^n$ contains 8 independent hermitian generators of which two (the trace, $i S_m^m$, and the $\Om$-trace, $\Om_m{}^n S_n^m$) commute with the other six. These six generators form one of $SO(4)$'s, whereas the two trace generators combine with 4 independent hermitian generators inside $S_{m n}$ and $S^{m n}$ to form another $SO(4)$. 
To see this, we first introduce parameters $\om^{m n}$, $\om_{m n}=(\om^{m n})^\ast$ and $\al_m^n=(\al_n^m)^\ast$ for the $R$-symmetry transformations, and define
\bea
\da_\om\phi = -\half(\om^{m n}S_{m n}-\om_{m n}S^{m n})\phi, \quad
\da_\al\phi = \al_m^n S_n^m\phi \ .
\eea
More explicitly,
\bea
\label{daomal2}
\da_\om\phi &=& \frac{1}{Z}\Big(\om^{m n}\Om_m{}^k q_{n k}-\om_{m n}\Om_k{}^m q^{n k}\Big)\phi \nn\\
\da_\al\phi &=& \al_m^n\Big[\Om_n{}^m\Big(-1+\frac{1}{Z}q^k q_k\Big)+\da_n^m\Big(\frac{1}{Z}\Om_k{}^l q^k q_l\Big)
-\frac{1}{Z}\Big(\Om_n{}^k q^m q_k+\Om_k{}^m q^k q_n\Big)\Big]\phi \ . \quad
\eea
The six 't Hooft matrices, $(\eta_\mfa)_{m n}$ and $(\wt\eta_\mfa)_{m n}$ (see Appendix~\ref{appTH}), form a basis in the space of antisymmetric 4 by 4 matrices, and so we write
\bea
\label{ometa}
\om^{m n} &=& \om_1\eta_1+\om_2\eta_2+\om_3\eta_3+\wt\om_1\wt\eta_1+\wt\om_2\wt\eta_2+\wt\om_3\wt\eta_3 \nn\\
\om_{m n} &=& \om_1^\ast\eta_1+\om_2^\ast\eta_2+\om_3^\ast\eta_3
+\wt\om_1^\ast\wt\eta_1+\wt\om_2^\ast\wt\eta_2+\wt\om_3^\ast\wt\eta_3 \ .
\eea
Conveniently enough, the products of 't Hooft matrices, $(\eta_a\wt\eta_b)_{m n}$, together with the unit matrix $I_4=\da_{m n}$ form a basis for all symmetric 4 by 4 matrices. This allows us to represent the hermitian matrix $\al_m^n$ as
\bea
\label{aleta}
\al_m^n &=& a_0 I_4+a_{1 1}\eta_1\wt\eta_1+a_{22}\eta_2\wt\eta_2+a_{33}\eta_3\wt\eta_3 \nn\\
&& +a_{12}\eta_1\wt\eta_2+a_{13}\eta_1\wt\eta_3+a_{21}\eta_2\wt\eta_1
+a_{23}\eta_2\wt\eta_3+a_{31}\eta_3\wt\eta_1+a_{32}\eta_3\wt\eta_2 \nn\\
&& +i(a_1\eta_1+a_2\eta_2+a_3\eta_3+\wt a_1\wt\eta_1+\wt a_2\wt\eta_2+\wt a_3\wt\eta_3) \ ,
\eea
where in the first, second and third lines we have, respectively, the diagonal, symmetric off-diagonal and antisymmetric matrices. All 16 coefficients $a$'s are real, whereas the 6 coefficients $\om$'s are complex. However, many of them drop out from $\da_\om\phi$ and $\da_\al\phi$ in (\ref{daomal2}) for symmetry reasons. With $\Om_m{}^n=(\eta_3)_{m n}$, we find that only
\bea
\om_1, \quad \om_2; \quad
a_0, \quad a_3, \quad
\wt a_1, \quad \wt a_2, \quad \wt a_3, \quad
a_{31}, \quad a_{32}, \quad a_{33}
\eea
contribute. This constitutes 12 real parameters, which matches the dimension of $SO(4)\times SO(4)$. More explicitly, substituting (\ref{ometa}) and (\ref{aleta}) into (\ref{daomal2}), we find that
\bea
\da_\om\phi &=& \Big(-\om_1\ov W_2-\om_1^\ast W_2+\om_2\ov W_1+\om_2^\ast W_1\Big)\phi \nn\\
\da_\al\phi &=& 2\Big[a_0 V+a_{3\mfa}B_\mfa-i(a_3 U+\wt a_\mfa A_\mfa)\Big]\phi \ ,
\eea
where $\mfa=1,2,3$ and we defined
\bea
\label{defUV}
&& W_\mfa=(\eta_\mfa)_{m n}\frac{1}{Z}q^{m n}, \quad
\ov W_\mfa=(\eta_\mfa)_{m n}\frac{1}{Z}q_{m n} \nn\\
&& U=-2+\frac{1}{Z}q^k q_k, \quad
V=\Om_m{}^n\frac{1}{Z}q^m q_n \nn\\
&& A_\mfa=(\eta_3\wt\eta_\mfa)_{m n}\frac{1}{Z}q^m q_n, \quad
B_\mfa=(\wt\eta_\mfa)_{m n}\frac{1}{Z}q^m q_n \ .
\eea
The hermiticity (or complex conjugation \cite{db}) properties of these generators are
\bea
U^\ast=U, \quad V^\ast=-V, \quad W_\mfa^\ast=-\ov W_\mfa, \quad
A_\mfa^\ast=A_\mfa, \quad B_\mfa^\ast=-B_\mfa \ .
\eea
Reorganizing them into the following four triplets of \emph{hermitian} generators
\bea
\ba[b]{rcl@{\qquad}rcl}
X_1 &=&\dst -\frac{1}{8}(W_2-\ov W_2)+\frac{i}{8}(W_1+\ov W_1), &
Y_1 &=&\dst \frac{1}{8}(W_2-\ov W_2)+\frac{i}{8}(W_1+\ov W_1) \\[10pt]
X_2 &=&\dst +\frac{1}{8}(W_1-\ov W_1)+\frac{i}{8}(W_2+\ov W_2), &
Y_2 &=&\dst \frac{1}{8}(W_1-\ov W_1)-\frac{i}{8}(W_2+\ov W_2) \\[10pt]
X_3 &=&\dst \qter(U-i V), &
Y_3 &=&\dst \qter(U+i V) \\[10pt]
R_\mfa &=&\dst -\half(A_\mfa+i B_\mfa), &
L_\mfa &=&\dst \half(A_\mfa-i B_\mfa) \ ,
\ea
\eea
we find that
\bea
[X_\mfa,X_\mfb]=i\eps_{\mfa \mfb \mfc}X_\mfc \ ,
\eea
and similarly for $Y_\mfa$, $R_\mfa$ and $L_\mfa$, whereas all other commutators vanish. This proves that the $R$-symmetry group of the mass-deformed theory is $SU(2)\times SU(2)\times SU(2)\times SU(2)$, which is the same as $SO(4)\times SO(4)$.

On another hand, it is also instructive to see how the $R$-symmetry transformations act on the scalars. Projecting (\ref{daomal2}) to find the corresponding variations of $A$ and $C_{m n}$, and using (\ref{phicomp}) together with (\ref{AandC}), we find after a little algebra that
\bea
\ba[b]{rcl@{\qquad}rcl}
\da_R X^3 &=& 4(-a_3 X^4-b_2 X^5+b_1 X^6), &
\da_R X^7 &=& 4(-\wt a_1 X^8-\wt a_2 X^9-\wt a_3 X^{10}) \\[5pt]
\da_R X^4 &=& 4(a_3 X^3-c_2 X^5+c_1 X^6), &
\da_R X^8 &=& 4(\wt a_1 X^7+a_{33}X^9-a_{32}X^{10}) \\[5pt]
\da_R X^5 &=& 4(b_2 X^3+c_2 X^4+a_0 X^6), &
\da_R X^9 &=& 4(\wt a_2 X^7-a_{33}X^8+a_{31}X^{10}) \\[5pt]
\da_R X^6 &=& 4(-b_1 X^3-c_1 X^4-a_0 X^5), &
\da_R X^{10} &=& 4(\wt a_3 X^7+a_{32}X^8-a_{31}X^9) \ ,
\ea
\eea
where $\da_R=\da_\om+\da_\al$, and we defined $\om_{1,2}=b_{1,2}+i c_{1,2}$. This clearly shows the $SO(4)\times SO(4)$ structure of the surviving $R$-symmetry transformations. 

\subsection{Hamiltonian as a quadratic form}

The on-shell Lagrangian, obtained by substituting (\ref{PsiMinus}) into (\ref{LwithM}), can be easily transformed into the superfield form,
\bea
\mc{L} &=& \half X^I(\Box-M^2)X^I-\frac{i}{2}\Psibar_{+}\Ga_{-}\frac{1}{2\p^{+}}(\Box-M^2)\Psi_{+} \nn\\
&=& \Abar(\Box-M^2)A+\qter C_{m n}(\Box-M^2)C^{m n}+\frac{i}{\sqrt2}\chi_m\frac{1}{\p^{+}}(\Box-M^2)\chi^m \nn\\
&=& -\frac{1}{8}\int d^4\ta d^4\ta \Big\{\phibar\frac{1}{\p^{+2}}(\Box-M^2)\phi\Big\} \ ,
\eea
where we used that
\bea
\label{d4d4}
\int d^4\ta d^4\tabar(\dots)=d^{[4]}d_{[4]}(\dots)_|
=\frac{1}{4!}\eps_{i j k l}d^{i j k l}\frac{1}{4!}\eps^{m n p q}d_{m n p q}(\dots)_| \ .
\eea
The LC Hamiltonian (defined with respect to the LC ``time'' $x^{+}$) is then~\footnote{
$\mc{H}$ should really be called the Hamiltonian density. The Hamiltonian is $H=\int dx^{-} dx^2\;\mc{H}$. We will work with $\mc{H}$ while freely dropping total $\p^{+}=-\p_{-}$ and $\p=\p_2$ derivatives, which is justified once we go back to $H$.
}
\bea
\mc{H}\equiv \frac{\da\mc{L}}{\da(\p^{-}\phi)}\p^{-}\phi-\mc{L}
=\frac{1}{8}\int d^4\ta d^4\ta\Big\{ \phibar\frac{1}{\p^{+2}}(\p^2-M^2)\phi\Big\} \ ,
\eea
where we used that $\Box=\p^2-2\p^{+}\p^{-}$. We claim that this can be rewritten as a quadratic form in the dynamical supersymmetry transformations,
\bea
\label{Hqfal}
\mc{H}=\frac{i}{16\sqrt2}\int d^4\ta d^4\ta\Big\{ (Q_m\phibar)\frac{1}{\p^{+}}(Q^m\phi) \Big\} \ .
\eea
The proof is the same as in \cite{abkr}, because, thanks to the following identity
\bea
2Z\Om_m{}^n q_n+\Om_m{}^n[q^k,q_{n k}]+\Om_k{}^n[q_{n m},q^k]=0 \ ,
\eea
we can simply integrate by parts with $Q_m$. Using (\ref{inout}), we then find that
\bea
\mc{H}=-\frac{i}{32\sqrt2}\int d^4\ta d^4\tabar \Big( \phibar\frac{1}{\p^{+}}\{Q_m,Q^m\}\phi \Big) \ ,
\eea
after which we use (\ref{QQPM}) to confirm the quadratic form property of the Hamiltonian.

\subsection{Dynamical Lorentz boost}

When $M=0$, conformal invariance allows the dynamical boost generator, $\mc{J}^{-}$, to be calculated by commuting the Hamiltonian shift generator, $\mc{P}^{-}$, with the kinematical generator of special conformal transformations, $K^{+}$ \cite{bbkr}. Once $M\neq0$, the conformal invariance is broken and $\mc{J}^{-}$ has to be derived from scratch. To do so, we start from the covariant form of the Lorentz transformations (\ref{Lorentz}).
Defining $\da_{\mc{J}^{-}}$ as the part of $\da_\la$ multiplied by $\la_{2-}$, we find that
\bea
\da_{\mc{J}^{-}} X^I &=& (x\p^{-}-x^{-}\p^{+})X^I \nn\\
\da_{\mc{J}^{-}}\Psi_{+} &=& (x\p^{-}-x^{-}\p^{+})\Psi_{+}+\frac{1}{2\p^{+}}(\p+M\Ga_2\Ga_{3456})\Psi_{+} \ ,
\eea
where as usual we set $x^{+}=0$ and substituted (\ref{PsiMinus}) for $\Psi_{-}$.
Reducing this to the variation of the superfield component fields, we obtain
\bea
\da_{\mc{J}^{-}}A &=& \Big(x\p^{-}-x^{-}\p\Big)A \nn\\
\da_{\mc{J}^{-}}\chi_m &=& \Big(x\p^{-}-x^{-}\p+\frac{\p}{2\p^{+}}\Big)\chi_m+\frac{i M}{2\p^{+}}\Om_m{}^n\chi_n \nn\\
\da_{\mc{J}^{-}}C_{m n} &=& \Big(x\p^{-}-x^{-}\p\Big)C_{m n} \ .
\eea
The superfield expression for the $M$-independent part of $\da_{\mc{J}^{-}}\phi$ follows easily by comparison with \cite{bbkr}. To find the corresponding expression for the $O(M)$ part, we observe that, with the $R$-symmetry generators $U$ and $V$ defined in (\ref{defUV}), we have
\bea
i U(A,\;\chi_m,\; C_{m n},\;\chi^m,\;\Abar) 
&=& (-2i A,\;-i\chi_m,\; 0,\; i\chi^m,\; 2i\Abar) \nn\\
V(A,\;\chi_m,\; C_{m n},\;\chi^m,\;\Abar) 
&=& (0,\;\Om_m{}^n\chi_n,\;\Om_m{}^k C_{k n}-\Om_n{}^k C_{k m},\;\Om_n{}^m\chi^n,0) \ ,
\eea
and therefore
\bea
i U V(A,\;\chi_m,\; C_{m n},\;\chi^m,\;\Abar) 
&=& (0,\; -i\Om_m{}^n\chi_n,\; 0,\; -i\Om_n{}^m\chi^n,\; 0) \ .
\eea
It then follows that
\bea
\label{Jminus}
\da_{\mc{J}^{-}}\phi=\Big(x\p^{-}-x^{-}\p+(\half\mc{N}-1)\frac{\p}{\p^{+}}\Big)\phi-\frac{i M}{2\p^{+}}U V\phi \ ,
\eea
where $\mc{N}=\ta^m\p_m+\ta_m\p^m=4+\frac{1}{Z}(q^m d_m+q_m d^m)$ \cite{bbkr} and
\bea
\p^{-}=\frac{1}{2\p^{+}}(\p^2-M^2) \ .
\eea
We thus have found the mass deformation of all the (non-conformal) dynamical generators of the (abelian) BLG theory. The kinematical (non-conformal) generators receive no $M$-dependent modifications, and are the same as in \cite{bbkr}. The conformal symmetries (dilatations, special conformal and superconformal) are, obviously, broken by the mass deformation. In the next section, we will discuss the non-abelian generalization of these results.

\newpage
\section{The full mass-deformed BLG theory}
\label{sec-fullBLG}

In this section, we extend the preceding results to the case when $f^{b c d}{}_a$ is non-zero. The mass deformation of the supersymmetry transformations has no terms with $M$ multiplying $f^{b c d}{}_a$, and so (\ref{phidsusy}) generalizes trivially. The $R$-symmetry generators are all kinematical, and receive no modifications from the $f$'s. The difficult part is to verify that the full LC superspace Hamiltonian can still be written as a quadratic form in the dynamical supersymmetry transformations. We will show that this is, indeed, the case.

\subsection{Supersymmetry transformations}

The complete supersymmetry transformations of the mass-deformed BLG theory are \cite{gsp,hll,lr}
\bea
\label{MBLGsusy}
\da_\ep X_a^I &=& i\epbar\Ga^I\Psi_a \nn\\
\da_\ep\Psi_a &=& D_\mu X_a^I\Ga^\mu\Ga^I\ep-\frac{1}{6}X_b^I X_c^J X_d^K f^{b c d}{}_a\Ga^{I J K}\ep
-M\Ga_{3456}\Ga^I X_a^I\ep \nn\\
\da_\ep\wt A_\mu{}^b{}_a &=& i\epbar\Ga_\mu\Ga^I X_c^I\Psi_d f^{c d b}{}_a \ ,
\eea
where $f^{b c d}{}_a$ are totally antisymmetric in the upper indices, and satisfy the Fundamental Identity \cite{gnp}
\bea
\label{FI}
f^{[e f g}{}_d f^{c]d b}{}_a=0 \ .
\eea
The commutator of these transformations closes into the translation, gauge transformation and an $M$-dependent $R$-symmetry transformation, plus terms proportional to the equations of motion. 
To derive the LC superspace transformation laws, we need to go through the following steps \cite{db}
\begin{itemize}
\item
fix the LC gauge $\wt A_{-}{}^b{}_a=0$;
\item
solve equations of motion for dependent field components ($\wt A_{+}{}^b{}_a$, $\wt A_2{}^b{}_a$ and $\Psi_{a-}$);
\item
modify supersymmetry transformations by adding compensating gauge transformations required to stay in the gauge;
\item
find the modified supersymmetry transformations of $A_a$, $C_{m n a}$ and $\chi_{m a}$; 
\item
guess and confirm the corresponding superfield transformation law.
\end{itemize}
As the equation of motion for $\wt A_\mu{}^b{}_a$ is not affected by the mass-deformation, the expressions for $\wt A_{+}{}^b{}_a$ and $\wt A_2{}^b{}_a$ remain the same as in \cite{db}. As there are no $M$-dependent corrections to $\da_\ep\wt A_\mu{}^b{}_a$, the parameters of the compensating gauge transformations $\wt\La^b{}_a$ are also the same as in \cite{db}. The equation of motion for $\Psi_a$ is
\bea
\Ga^\mu D_\mu\Psi_a+\half\Ga^{I J}\Psi_b X_c^I X_d^J f^{b c d}{}_a+M\Ga_{3456}\Psi_a=0 \ ,
\eea
and so the expression for $\Psi_{a-}$ is modified. However, as the $M$-dependent term comes without $f^{b c d}{}_a$, its effect on the supersymmetry transformation of $\phi_a$ is exactly the same as in the $f^{b c d}{}_a=0$ case considered in Section \ref{sec-abelianBLG}.~\footnote{
Similar analysis shows that our result (\ref{Jminus}) captures all the $M$-dependence of the dynamical Lorentz boost, $\da_{\mc{J}^{-}}\phi_a$. Note that $\p^{-}\phi_a$ there receives an additional $O(M)$ correction, which can be deduced from the full Hamiltonian.
}
Therefore, combining (\ref{phidsusy}) with the results of~\cite{db}, we conclude that the full LC superspace transformation laws in the mass-deformed BLG theory are
\bea
\label{kdfsusy}
\da_{\ep_{-}}\phi_a &=& \sqrt2(\al^m q_m-\al_m q^m)\phi_a \nn\\
\da_{\ep_{+}}\phi_a &=& (\beta^m Q_m-\beta_m Q^m)\phi_a+i\beta_m W_a^m
+\frac{d^{[4]}}{2\p^{+2}}(i\beta^m W_{m a}) \ ,
\eea
where $Q$'s are given in (\ref{QQbar}), and 
\bea
W_a^m &=& -\frac{i}{3\sqrt2}\eps^{m n k l}\frac{1}{\p^{+}}\Big(\p^{+}\phi_b\cdot\frac{1}{\p^{+}}
(\p^{+}\phi_c\cdot q_{n k l}\phi_d+3\p^{+}q_n\phi_c\cdot q_{k l}\phi_d)\Big)f^{b c d}{}_a \nn\\
W_{m a} &=& -\frac{i}{3\sqrt2}\eps_{m n k l}\frac{1}{\p^{+}}\Big(\p^{+}\phibar_b\cdot\frac{1}{\p^{+}}
(\p^{+}\phibar_c\cdot q^{n k l}\phibar_d+3\p^{+}q^n\phibar_c\cdot q^{k l}\phibar_d)\Big)f^{b c d}{}_a \ .
\eea
Note that we chose the form of $W$'s that involves $q$'s instead of $d$'s (see (7.43) in \cite{db}). As we will see, with this choice the proof of the quadratic form property of the LC superspace Hamiltonian simplifies tremendously (cf.~\cite{bbkr}).

\subsection{The Lagrangian and the LC Hamiltonian}

In order to write down the Lagrangian invariant under the supersymmetry transformations (\ref{MBLGsusy}), we need a metric $h_{a b}$ for raising and lowering the gauge indices. Requiring that the resulting $f^{a b c d}\equiv f^{a b c}{}_e h^{e d}$ is totally antisymmetric, the Lagrangian is given by
\bea
\mc{L} &=& \mc{L}_{BLG}+\mc{L}_M \nn\\
\mc{L}_{BLG} &=& -\half(D_\mu X_a^I)(D^\mu X_a^I)+\frac{i}{2}\Psibar_a\Ga^\mu D_\mu\Psi_a
+\frac{i}{4}\Psibar_b\Ga_{I J} X_c^I X_d^J\Psi_a f^{b c d a} \nn\\
&&+\half\eps^{\mu\nu\la}A_{\mu a b}(\p_\nu\wt A_\la{}^{a b}+\frac{2}{3}\wt A_\nu{}^a{}_c\wt A_\la{}^{c b})
-\frac{1}{12}f^{a b c d}f^{e f g}{}_d X_a^I X_b^J X_c^K X_e^I X_f^J X_g^K \nn\\
\mc{L}_M &=& -\frac{M^2}{2}X_a^I X_a^I+\frac{i}{2}M\Psibar_a\Ga_{3456}\Psi_a
-4M f^{a b c d}(X_a^3 X_b^4 X_c^5 X_d^6+X_a^7 X_b^8 X_c^9 X_d^{10}) \ ,
\eea
where $D_\mu\Psi_a=\p_\mu\Psi_a-\wt A_\mu{}^b{}_a\Psi_b$ with
$\wt A_\mu{}^b{}_a=f^{c d b}{}_a A_{\mu c d}$. 
In the LC gauge, with all the dependent fields substituted into the Lagrangian (as in \cite{bn} but in the conventions of \cite{db}), the $\Psi$-independent part of the Lagrangian reduces to $\mc{L}_X=-X_a^I\p^{+}\p^{-}X_a^I-\mc{H}_X$ where
\bea
\label{HX}
\mc{H}_{X} &=& -\half X_a^I(\p^2-M^2)X_a^I-f^{a b c d}(X_a^I\p X_b^I)\frac{1}{\p^{+}}(X_c^J\p^{+}X_d^J) \nn\\
&& +\half f^{a b c d}f^{a b^\rr c^\rr d^\rr}(X_b^I X_{b^\rr}^I)
\cdot\frac{1}{\p^{+}}(X_c^J\p^{+} X_d^J)
\cdot\frac{1}{\p^{+}}(X_{c^\rr}^K X_{d^\rr}^K) \nn\\
&&+\frac{1}{12}f^{a b c d}f^{a b^\rr c^\rr d^\rr}(X_b^I X_{b^\rr}^I)(X_c^J X_{c^\rr}^J)(X_d^J X_{d^\rr}^J) \nn\\[5pt]
&& +4M f^{a b c d}(X_a^3 X_b^4 X_c^5 X_d^6+X_a^7 X_b^8 X_c^9 X_d^{10}) \ .
\eea
The first three lines are $SO(8)$ invariant and can be rewritten in terms of $A$'s and $C$'s using 
\bea
X^I X^{\rr I}=A\Abar^\rr+\Abar A^\rr+\half C_{m n} C^{\rr m n} \ .
\eea
The $O(M)$ part of the LC Hamiltonian breaks $SO(8)$ down to $SO(4)\times SO(4)$. Its form in terms of $A$'s and $C$'s is given in equation (\ref{HMX}) of Appendix~\ref{appH1}.

\subsection{Hamiltonian as a quadratic form}

According to (\ref{kdfsusy}), we have 
$\da_{\ep_{+}}\phi_a = \da_{\betabar\mc{Q}}\phi_a+\da_{\beta\ov{\mc{Q}}}\phi_a$ where
\bea
\da_{\betabar\mc{Q}}\phi_a &=& \beta_m(-Q^m\phi_a+i W_a^m), \quad
\da_{\beta\ov{\mc{Q}}}\phi_a=\frac{d^{[4]}}{2\p^{+2}}(\da_{\betabar\mc{Q}}\phi_a)^\ast \nn\\
(\da_{\betabar\mc{Q}}\phi_a)^\ast &=& \beta^m(Q_m\phibar_a+i W_{m a}) \ .
\eea
We therefore expect that the quadratic form property of the Hamiltonian with $f^{a b c d}=0$, see (\ref{Hqfal}), generalizes to
\bea
\label{Hqf}
\mc{H}=\frac{i}{16\sqrt2}\int d^4\ta d^4\tabar\ (Q_m\phibar_a+i W_{m a})\frac{1}{\p^{+}}(Q^m\phi_a-i W_a^m) \ .
\eea
Expanding in powers of $f$'s, we have $\mc{H}=\mc{H}^{(0)}+\mc{H}^{(1)}+\mc{H}^{(2)}$, where
\bea
\mc{H}^{(0)} &=& \frac{i}{16\sqrt2}\int d^4\ta d^4\tabar\ (Q_m\phibar_a)\frac{1}{\p^{+}}(Q^m\phi_a)
\nn\\
\mc{H}^{(1)} &=& -\frac{1}{16\sqrt2}\int d^4\ta d^4\tabar\ (Q^m\phi_a)\frac{1}{\p^{+}}W_{m a} +c.c.
\nn\\
\mc{H}^{(2)} &=& \frac{i}{16\sqrt2}\int d^4\ta d^4\tabar\ W_{m a}\frac{1}{\p^{+}} W_a^m \ .
\eea
We have already verified that $\mc{H}^{(0)}$ reproduces the corresponding part in (\ref{HX}). In Appendices \ref{appH1} and \ref{appH2}, we show that the same is true for the $O(f)$ and $O(f^2)$ parts. Kinematical supersymmetry then guarantees that the $\Psi$-dependent part of $\mc{H}$ is reproduced correctly as well. The quadratic form property of the LC superspace Hamiltonian is therefore rigorously established.

\section{Summary and discussion}

In this paper, we have analyzed the mass-deformed BLG theory \cite{gsp,hll} from the LC superspace point of view \cite{bln1}. We found that the mass deformation is treated as an interaction in the sense that the (surviving) kinematical symmetry generators \cite{bbkr}, including the kinematical supersymmetry generators $q$'s, are not modified by it.~\footnote{
Other kinematical generators, including the conformal ones and part of the $SO(8)$ $R$-symmetry, cease to represent symmetries when the mass $M$ is introduced.
}
The (surviving) dynamical symmetry generators, the dynamical supersymmetries $\mc{Q}$'s, the Hamiltonian shift $\mc{P}^{-}$, and the dynamical Lorentz boost $\mc{J}^{-}$, all receive $M$-dependent corrections. 

The modification of the $\mc{Q}$'s is the simplest, but non-trivial: it is linear in $M$, independent of $f^{b c d}{}_a$, proportional to $\Om_m{}^n$ and cubic in the $q$'s. The matrix $\Om$ carries $SU(4)$ indices and specifies a quaternionic direction in the algebra of $SO(4)\subset SU(4)$. It plays the key role in reducing the $SO(8)$ $R$-symmetry of the BLG theory down to $SO(4)\times SO(4)$. 

The $\mc{P}^{-}$ can be determined either from the commutator of two $Q$'s or as a functional derivative of the LC superspace Hamiltonian $H$ \cite{bbkr}. The resulting expression is complicated, and has both $O(M)$ and $O(M^2)$ parts. However, $H$ itself is extremely simple, being given by the quadratic form (\ref{Hqf}). 

The $M$-independent part of $\mc{J}^{-}$ can be derived by commuting $\mc{P}^{-}$ with the kinematical special conformal generator $K^{+}$ \cite{bbkr}. Most of its $M$-dependence then comes from adjusting the value of $\p^{-}$, encoded in the Hamiltonian. Its remaining $O(M)$ part, see (\ref{Jminus}), turns out to be given by a single term quadratic in the surviving $R$-symmetry generators.

The quadratic form property of the LC superspace Hamiltonian $H$ has been established via an explicit calculation. We extended the proof of \cite{bbkr} to the quadratic order in $f^{a b c d}$, as well as proved that the same quadratic form correctly describes the $M$-dependent parts of~$H$. Still lacking, however, is the fundamental understanding of this property (first observed in \cite{abkr}). It appears to be rooted into maximal supersymmetry, which in turn imposes the reality (``inside-out'') constraint (\ref{inout}) on the superfield $\phi$ \cite{bln1} and leads to the quadratic form property at the $O(f^0)$ level. Presumably, the preservation of this property while turning on the structure constants $f$'s can be attributed to analyticity of extended supersymmetry \cite{gio}. 

The mass deformation of the BLG theory serves as a supersymmetry preserving IR regulator, and therefore we expect that the analysis performed in this paper should be useful for studying quantum properties of the BLG theory in LC superspace. 

It would also be very interesting to understand the LC superspace formulation of theories with less-than-maximal supersymmetry (such as \cite{abjm,bl4}). In particular, to study deviations from the quadratic form property of the LC superspace Hamiltonians in these cases.

\vspace{20pt}
\noindent{\bf\large Acknowledgements}
\vspace{10pt}

\noindent
I thank Jon Bagger, Lars Brink and Pierre Ramond for helpful discussions, and Sung-Soo Kim for informing me of identity (\ref{fourCid}). This research was supported by the Department of Energy Grant No. DE-FG02-97ER41029.

\vspace{10pt}

\appendix

\section{Gamma matrices and 't Hooft symbols}
\label{appTH}

The representation of the $d=11$ gamma matrices that we use is as follows \cite{db}
\bea
\ba[b]{rcl@{\qquad}rcl}
\Ga_0 &=& -i(I_2\otimes\si_1)\otimes(I_4\otimes I_2), &
\Ga_5 &=& -i(I_2\otimes\si_3)\otimes(\eta_1\otimes\si_1) \nn\\[3pt]
\Ga_1 &=& i(\si_3\otimes i\si_2)\otimes(I_4\otimes I_2), & 
\Ga_6 &=& -i(I_2\otimes\si_3)\otimes(\eta_2\otimes\si_1) \nn\\[3pt]
\Ga_2 &=& (I_2\otimes\si_3)\otimes(I_4\otimes\si_3), &
\Ga_7 &=& -i(I_2\otimes\si_3)\otimes(\eta_3\otimes\si_1) \nn\\[3pt]
\Ga_3 &=& i(\si_1\otimes i\si_2)\otimes(I_4\otimes I_2), &
\Ga_8 &=& -(I_2\otimes\si_3)\otimes(\wt\eta_1\otimes i\si_2) \nn\\[3pt]
\Ga_4 &=& i(\si_2\otimes i\si_2)\otimes(I_4\otimes I_2), &
\Ga_9 &=& -(I_2\otimes\si_3)\otimes(\wt\eta_2\otimes i\si_2) \nn\\[3pt]
&& &
\Ga_{10} &=& -(I_2\otimes\si_3)\otimes(\wt\eta_3\otimes i\si_2) \ ,
\ea
\eea
where the Pauli matrices $\si_\mfa$, $\mfa=1,2,3$, are standard
\bea
I_2=\bpm 1 & 0 \\ 0 & 1 \epm, \quad
\si_1=\bpm 0 & 1 \\ 1 & 0 \epm, \quad
i\si_2=\bpm 0 & 1 \\ -1 & 0 \epm, \quad
\si_3=\bpm 1 & 0 \\ 0 & -1 \epm \ ,
\eea
and the 't Hooft symbols $\eta_{\mfa m n}$ and $\wt\eta_{\mfa m n}$ are given by
\bea
\ba[b]{l@{\qquad}l}
\eta_1=+\si_1\otimes i\si_2=\bpm & {\bma & + \\ + & \ema} \\ {\bma & - \\ - & \ema} \epm
& 
\wt\eta_1=-i\si_2\otimes\si_1=\bpm & {\bma & - \\ + & \ema} \\ {\bma & - \\ + & \ema} \epm
\\[30pt]
\eta_2=-\si_3\otimes i\si_2=\bpm & {\bma - & \\ & + \ema} \\ {\bma + & \\ & - \ema} \epm
&
\wt\eta_2=-I_2\otimes i\si_2=\bpm & {\bma - & \\ & - \ema} \\ {\bma + & \\ & + \ema} \epm
\\[30pt]
\eta_3=+i\si_2\otimes I_2=\bpm {\bma & + \\ - & \ema} & \\ & {\bma & + \\ - & \ema} \epm
&
\wt\eta_3=+i\si_2\otimes\si_3=\bpm {\bma & + \\ - & \ema} & \\ & {\bma & - \\ + & \ema} \epm \ .
\ea
\eea
They satisfy $\si_\mfa \si_\mfb=I_2\da_{\mfa\mfb}+i\eps_{\mfa\mfb\mfc}\si_{\mfc}$ and~\footnote{
The $(I_4,\eta_\mfa)$ and $(I_4,\wt\eta_\mfa)$ are two representations of quaternions as $SO(4)$ rotation matrices corresponding to left- and right-multiplication of quaternions, respectively (see e.g. \cite{mebius}).
}
\bea
\eta_\mfa\eta_\mfb=-I_4\da_{\mfa\mfb}-\eps_{\mfa\mfb\mfc}\eta_{\mfc}, \quad
\eta_\mfa\wt\eta_\mfb=\wt\eta_\mfb\eta_\mfa, \quad
\wt\eta_\mfa\wt\eta_\mfb=-I_4\da_{\mfa\mfb}-\eps_{\mfa\mfb\mfc}\wt\eta_{\mfc} \ .
\eea
Note that $\eta_1\eta_2\eta_3=\wt\eta_1\wt\eta_2\wt\eta_3=I_4$. The only independent products, therefore, are
\bea
&&
\eta_1\wt\eta_1=\bpm {\bma + & \\ & - \ema} & \\ & {\bma - & \\ & + \ema} \epm, \quad
\eta_1\wt\eta_2=\bpm {\bma & + \\ + & \ema} & \\ & {\bma & + \\ + & \ema} \epm, \quad
\eta_1\wt\eta_3=\bpm & {\bma + & \\ & - \ema} \\ {\bma + & \\ & - \ema} & \epm
\nn\\
&&
\eta_2\wt\eta_1=\bpm {\bma & + \\ + & \ema} & \\ & {\bma & - \\ - & \ema} \epm, \quad
\eta_2\wt\eta_2=\bpm {\bma - & \\ & + \ema} & \\ & {\bma - & \\ & + \ema} \epm, \quad
\eta_2\wt\eta_3=\bpm & {\bma & + \\ + & \ema} \\ {\bma & + \\ + & \ema} & \epm
\nn\\
&&
\eta_3\wt\eta_1=\bpm & {\bma + & \\ & + \ema} \\ {\bma + & \\ & + \ema} & \epm, \quad
\eta_3\wt\eta_2=\bpm & {\bma & - \\ + & \ema} \\ {\bma & + \\ - & \ema} & \epm, \quad
\eta_3\wt\eta_3=\bpm {\bma - & \\ & - \ema} & \\ & {\bma + & \\ & + \ema} \epm \ . \hspace{35pt}
\eea
The $d=11$ charge conjugation matrix is
\bea
C=i(i\si_2\otimes\si_3)\otimes(I_4\otimes\si_1) \ ,
\eea
and it is used to define the conjugated spinors $\epbar=\ep^T C$ and $\Psibar=\Psi^T C$. We also observe that
\bea
\Ga_{012} &=& -(\si_3\otimes I_2)\otimes(I_4\otimes\si_3) \nn\\
\Ga_{3456} &=& i(\si_3\otimes I_2)\otimes(\eta_3\otimes I_2) \nn\\
\Ga_{789(10)} &=& i(I_2\otimes I_2)\otimes(\eta_3\otimes\si_3) \ ,
\eea
from which $\Ga_{789(10)}=-\Ga_{3456}\Ga_{012}$ follows. Finally, our conventions \cite{db} are such that
\bea
\left\{\mc{M}_4\otimes\bpm a^m{}_n & b^{m n} \\ c_{m n} & d_m{}^n \epm\right\}
\bpm \psi^n \\ \psi_n \epm
=\bpm \mc{M}_4(a^m{}_n\psi^n+b^{m n}\psi_n) \\ \mc{M}_4(c_{m n}\psi^n+d_m{}^n\psi_n) \epm \ ,
\eea
where the 4 by 4 matrix $\mc{M}_4$ acts on the (implicit) spinor indices of $\psi$'s.

\section{Useful identities}
\label{appID}

The self-dual tensor $C_{m n}$, satisfying
\bea
(C_{m n})^\ast=C^{m n}=\half\eps^{m n k l}C_{k l} \ ,
\eea
enjoys many interesting identities. The basic identity that we need is
\bea
\label{twoCid}
(C_{i k},C^{j k})+(C^{j k},C_{i k})=\half (C_1,C_1)\da_i^j \ ,
\eea
where $(C_1,C_1)\equiv(C_{m n},C^{m n})=(C^{m n},C_{m n})$. Using the shorthand notation $C_{12}\equiv C_{i_1 i_2}$, etc., we find that
\bea
(\un{C_{12}},\un{C^{23}},C_{34},C^{41})+(C^{23},C_{12},C_{34},C^{41})=\half(C_1,C_1,C_2,C_2) \ ,
\eea
where we underlined the two $C$'s to which the identity (\ref{twoCid}) is applied.
Noting that
\bea
\label{cycleC4}
&& \Big[(\un{C_{12}},\un{C^{23}},C_{34},C^{41})+(C^{23},C_{12},C_{34},C^{41})\Big] \nn\\
&-&\Big[(\un{C^{23}},C_{12},\un{C_{34}},C^{41})+(C_{34},C_{12},C^{23},C^{41})\Big] \nn\\
&+&\Big[(C_{34},\un{C_{12}},\un{C^{23}},C^{41})+(C_{34},C^{23},C_{12},C^{41})\Big] = 2(C_{12},C^{23},C_{34},C^{41}) \ ,
\eea
we deduce the following identity
\bea
\label{fourCid}
(C_{12},C^{23},C_{34},C^{41})=\qter\Big\{
(C_1,C_1,C_2,C_2)-(C_1,C_2,C_1,C_2)+(C_1,C_2,C_2,C_1)\Big\} \ .
\eea
Denoting the LHS of (\ref{cycleC4}) as ``$(12/23)-(23/34)+(12/23)$,'' we find that the corresponding sequence needed to similarly reduce $(C_{12},C^{23},C_{34},C^{45},C_{56},C^{61})$ is
\bea
&& (12/23)-(23/34)+(34/45)-(12/23)+(23/34)-(12/23)+(56/61) \nn\\
&=& (C_{12},C^{23},C_{34},C^{45},C_{56},C^{61})
+(C^{45},C_{34},C^{23},C_{12},C^{61},C_{56}) \ ,
\eea
which yields
\bea
(C_{12},C^{23},C_{34},C^{45},C_{56},C^{61})+c.c. &=& \half\Big\{
(C_5,C_5,C_{12},C^{23},C_{34},C^{41}) \nn\\
&&\hspace{-100pt}
-(C_5,C_{12},C_5,C^{23},C_{34},C^{41})
+(C_5,C_{12},C^{23},C_5,C_{34},C^{41}) \nn\\
&&\hspace{-100pt}
-(C_{12},C_5,C_5,C^{23},C^{41},C_{34})
+(C_{12},C_5,C^{23},C_5,C^{41},C_{34}) \nn\\
&&\hspace{-100pt}
-(C_{12},C^{23},C_5,C_5,C^{41},C_{34})
+(C_{12},C^{23},C_{34},C^{41},C_5,C_5)\Big\} \ .
\eea
Now the identity (\ref{fourCid}) can be applied and we find 21 terms on the RHS, of which 6 terms cancel upon relabelling of indices. The remaining 15 terms combine in a particularly nice way if we order the $C$'s as follows
\bea
\label{sixCid}
&&(C_{12},C^{61},C_{56},C^{23},C_{34},C^{45})+c.c.=-\frac{1}{8}\Big\{
522(511+115-151) \nn\\
&&+(512-521)(512+125-152)
+(112-121)(255+552-525) \Big\} \ ,
\eea
where $522511\equiv(C_5,C_2,C_2,C_5,C_1,C_1)$, etc.

\section{The linear in $f^{a b c d}$ part of the Hamiltonian}
\label{appH1}

The $O(f)$ part of the quadratic form Hamiltonian (\ref{Hqf}) is given by
\bea
\mc{H}^{(1)} = -\frac{1}{16\sqrt2} d^{[4]} \Big\{ Q^m d_{[4]}\phi_a\cdot\frac{1}{\p^{+}}W_{m a} \Big\}_|+c.c. \ ,
\eea
where we used (\ref{d4d4}) and the fact that $d_n W_{m a}=0$. For the $M$-independent part, we have
\bea
\mc{H}_{BLG}^{(1)} &=& -\frac{1}{48}f^{a b c d}\eps_{m n k l} \times \nn\\
&&\times d^{[4]}\Big\{
\frac{\p}{\p^{+}}q^m\phibar_a\cdot\p^{+}\phibar_b\cdot\frac{1}{\p^{+}}
(\p^{+}\phibar_c\cdot q^{n k l}\phibar_d+3\p^{+}q^n\phibar_c\cdot q^{k l}\phibar_d)\Big\}_| +c.c. \qquad
\eea
For the ``$C$-only'' part of the projection, only the term with ``3'' contributes, and there is only one way in which the four derivatives in $d^{[4]}=\frac{1}{4!}\eps_{r s t u}d^{r s t u}$ should be distributed among the four $\phibar$'s. We thus immediately find that~\footnote{
It is convenient to keep the gauge indices, together with $f^{a b c d}$, implicit. The antisymmetry of $f^{a b c d}$ translates into the fermionic-like behavior of the four objects separated by the central dots.
}
\bea
\mc{H}_{BLG|C^4}^{(1)}=\half\frac{\p}{\p^{+}}C^{m n}\cdot\p^{+}C_{m i}\cdot\frac{1}{\p^{+}}
(\p^{+}C^{i j}\cdot C_{n j})+c.c.
\eea
This corresponds to equation (H.10) in \cite{bbkr}, which there took much more effort to derive. Our simplified derivation is the consequence of using the expression of $W_{m a}$ in terms of $q$'s (rather than in terms of $d$'s). Using the identity (\ref{fourCid}) and the antisymmetry of $f^{a b c d}$, we find 
\bea
\mc{H}_{BLG|C^4}^{(1)}
&=& \qter\Big\{\frac{\p}{\p^{+}}C_1\cdot\p^{+}C_{2}\cdot\frac{1}{\p^{+}}(\p^{+}C_2\cdot C_1-\p^{+}C_1\cdot C_2) 
+\frac{\p}{\p^{+}}C_1\cdot\p^{+}C_1\cdot\frac{1}{\p^{+}}(\p^{+}C_2\cdot C_2)\Big\} \nn\\
&=& \qter\frac{\p}{\p^{+}}C_1\cdot\p^{+}\Big[C_1\cdot\frac{1}{\p^{+}}(\p^{+}C_2\cdot C_2)\Big]
= -\qter \p C_1\cdot C_1\cdot\frac{1}{\p^{+}}(\p^{+}C_2\cdot C_2) \ ,
\eea
which agrees with (\ref{HX}). We verified that $\mc{H}_{BLG|A^2 C^2}^{(1)}$ and $\mc{H}_{BLG|A^4}^{(1)}$ match with (\ref{HX}) as well, together with $\mc{H}_{BLG|C^4}^{(1)}$ giving
\bea
\mc{H}_{BLG|X}^{(1)}=-(A\cdot\p\Abar+\Abar\cdot\p A+\half C_1\cdot\p C_1)\cdot\frac{1}{\p^{+}}
(A\cdot\p^{+}\Abar+\Abar\cdot\p^{+}A+\half C_2\cdot\p^{+}C_2) \ . \qquad
\eea
Turning to the $M$-dependent part of $\mc{H}^{(1)}$, we find that
\bea
\label{HqfM1}
\mc{H}_M^{(1)} &=& -\frac{M}{48\sqrt2}f^{a b c d}\eps_{m n k l} \times \nn\\
&&\times d^{[4]}\Big\{
\frac{1}{\p^{+2}}V^m\phibar_a\cdot\p^{+}\phibar_b\cdot\frac{1}{\p^{+}}
(\p^{+}\phibar_c\cdot q^{n k l}\phibar_d+3\p^{+}q^n\phibar_c\cdot q^{k l}\phibar_d)\Big\}_| +c.c. \ , \qquad
\eea
where
\bea
V^m\equiv Z\Om_n{}^m q^n+\Om_n{}^m q_k q^{n k}+\Om_n{}^k q_k q^{m n} \ ,
\eea
which has the following projections
\bea
d^r V^m\phibar_|=-Z\Om_p{}^r d^{p m}\phibar_|, \quad
\eps_{r s t u}d^{r s t}V^m\phibar_|=-12Z\Om_u{}^m\p^{+2}\phi_| \ .
\eea
Acting with $d^{[4]}=\frac{1}{4!}\eps_{r s t u}d^{r s t u}$, we easily find that the ``$A$-only'' part of $\mc{H}_M^{(1)}$ vanishes,
\bea
\label{HMA4}
\boxed{
\mc{H}_{M|A^4}^{(1)}=0 
} \ ,
\eea 
as it is proportional to $\Om_m{}^m=0$. For the part with two $A$'s and two $C$'s, we find
\bea
\mc{H}_{M|A^2 C^2}^{(1)}=\frac{i}{2}M\Om_n{}^m\Big\{
\frac{1}{\p^{+}}C^{n k}\cdot\p^{+}C_{m k}\cdot\frac{1}{\p^{+}}(\Abar\cdot\p^{+}A)
+A\cdot\Abar\cdot\frac{1}{\p^{+}}(\p^{+}C^{n k}\cdot C_{m k}) \nn\\
-\frac{1}{\p^{+}}C^{n k}\cdot\Abar\cdot\frac{1}{\p^{+}}(\p^{+}C_{m k}\cdot\p^{+}A)
+\frac{1}{\p^{+}}C^{n k}\cdot\Abar\cdot\frac{1}{\p^{+}}(\p^{+2}A\cdot C_{m k})\Big\}+c.c. \qquad
\eea
Using the total antisymmetry of $f^{a b c d}$, complex conjugation rules
\bea
(A)^\ast=\Abar, \quad (C_{m n})^\ast=C^{m n}, \quad (\Om_m{}^n)^\ast=-\Om_n{}^m \ ,
\eea
and the following identity
\bea
\Om_n{}^m(C^{n i},C_{m i})=-\Om_n{}^m(C_{m i},C^{n i}) \ ,
\eea
which follows from (\ref{twoCid}) and $\Om_m{}^m=0$, it is straightforward to prove that 
\bea
\label{HMA2C2}
\boxed{
\mc{H}_{M|A^2 C^2}^{(1)}=i M\Om_n{}^m(C^{n k}\cdot C_{m k}\cdot A\cdot\Abar)
} \ .
\eea
For the ``$C$-only'' part, we find
\bea
\mc{H}_{M|C^4}^{(1)}=-\frac{i}{2}M\Om_k{}^m(\frac{1}{\p^{+}}C^{k n}\cdot\p^{+}C_{m i}\cdot
\frac{1}{\p^{+}}(\p^{+}C^{i j}\cdot C_{n j}))+c.c.
\eea
In order to simplify this, it helps to split $i\Om_k{}^m C^{k n}$ into two parts, symmetric in $m n$ and antisymmetric in $m n$,
\bea
S^{m n}=\half(V^{m n}+V^{n m}), \quad
A^{m n}=\half(V^{m n}-V^{n m}); \quad
V^{m n}\equiv i\Om_k{}^m C^{k n} \ .
\eea
Note that $(A^{m n})^\ast=-\half\eps_{m n k l}A^{k l}$. Using (\ref{twoCid}), we find that
\bea
\frac{1}{\p^{+}}A^{m n}\cdot\p^{+}C_{m i}\cdot\frac{1}{\p^{+}}(\p^{+}C^{i j}\cdot C_{n j})
=-\qter A^{m n}\cdot C_{m n}\cdot\frac{1}{\p^{+}}(\p^{+}C_{i j}\cdot C^{i j}) \ .
\eea
As this expression is purely imaginary, $A^{m n}$ does not contribute to $\mc{H}_{M|C^4}^{(1)}$. Turning to the contribution of $S^{m n}$, we note that (\ref{twoCid}) and $S^{m n}C_{m n}=0$ imply that 
$(S^{m n}, C_{m i}, C^{i j}, C_{n j})$
is totally antisymmetric in the last three arguments. It is then straightforward to show that
\bea
\frac{1}{\p^{+}}S^{m n}\cdot\p^{+}C_{m i}\cdot\frac{1}{\p^{+}}(\p^{+}C^{i j}\cdot C_{n j})
=\frac{1}{6}(S^{m n}\cdot C^{i j}\cdot C_{m i}\cdot C_{n j}) \ ,
\eea
which is real. As a result,
\bea
\label{HMC4}
\boxed{
\mc{H}_{M|C^4}^{(1)}=-\frac{i}{6}M\Om_k{}^m(C^{k n}\cdot C^{i j}\cdot C_{m i}\cdot C_{n j})
} \ .
\eea
Finally, using that $\Om_m{}^n=(\eta_3)_{m n}$ and the expressions (\ref{AandC}), we find
\bea
\Om_n{}^m(C^{n k}\cdot C_{m k}\cdot A\cdot\Abar)
&=& \Om_n{}^m(C_{m k}\cdot C^{k n}\cdot A\cdot\Abar) \nn\\
&=& 2!Tr(\eta_3\eta_1\eta_2)\half(X^5\cdot X^6\cdot A\cdot\Abar) \nn\\
&=& 4(X^5\cdot X^6\cdot A\cdot\Abar) \nn\\
&=& -4i(X^5\cdot X^6\cdot X^3\cdot X^4) \ ,
\nn\\
\Om_k{}^m(C^{k n}\cdot C^{i j}\cdot C_{m i}\cdot C_{n j}) 
&=& \Om_k{}^m(C_{m i}\cdot C^{i j}\cdot C_{j n}\cdot C^{n k}) \nn\\
&=& 4! Tr(\eta_3\eta_3\wt\eta_1\wt\eta_2\wt\eta_3)\qter(X^7\cdot(-i X^8)\cdot(i X^9)\cdot(-i X^{10})) \nn\\
&=& 4!i(X^7\cdot X^8\cdot X^9\cdot X^{10}) \ ,
\eea
so that the sum of (\ref{HMA4}), (\ref{HMA2C2}) and (\ref{HMC4}) gives
\bea
\label{HMX}
\mc{H}_{M|X}^{(1)} &=&
i M\Om_n{}^m(C^{n k}\cdot C_{m k}\cdot A\cdot\Abar)
-\frac{i}{6}M\Om_k{}^m(C^{k n}\cdot C^{i j}\cdot C_{m i}\cdot C_{n j}) \nn\\
&=& 4M(X^3\cdot X^4\cdot X^5\cdot X^6+X^7\cdot X^8\cdot X^9\cdot X^{10}) \ ,
\eea
in agreement with (\ref{HX}).

\section{The quadratic in $f^{a b c d}$ part of the Hamiltonian}
\label{appH2}

The part of the quadratic form Hamiltonian (\ref{Hqf}) quadratic in $f^{a b c d}$ is
\bea
\mc{H}^{(2)}=\frac{i}{16\sqrt2}\int d^4\ta d^4\tabar\ W_{m a}\frac{1}{\p^{+}}W_a^m \ .
\eea
More explicitly, 
\bea
\label{Hqf2}
\mc{H}^{(2)} &=& \frac{i}{16\sqrt2\cdot 18}\eps_{m n p q}\eps^{m r s t}f^{a b c d}f^{a b^\rr c^\rr d^\rr}
d^{[4]} d_{[4]} \nn\\
&\times&\Big\{
\p^{+}\phibar_b\cdot\frac{1}{\p^{+}}
(\p^{+}\phibar_c\cdot q^{n p q}\phibar_d+3\p^{+}q^n\phibar_c\cdot q^{p q}\phibar_d) \nn\\
&&\cdot\frac{1}{\p^{+3}}(\p^{+}\phi_{b^\rr}\cdot\frac{1}{\p^{+}}
(\p^{+}\phi_{c^\rr}\cdot q_{r s t}\phi_{d^\rr}+3\p^{+}q_r\phi_{c^\rr}\cdot q_{s t}\phi_{d^\rr}))\Big\}_| \ .
\eea
Thanks to our choice of writing $W$'s in terms of $q$'s, $d_{[4]}=\frac{1}{4!}\eps^{i j k l}d_{i j k l}$ goes through the second line (since $d_m\phibar=0$). Acting on the third line, $d_{[4]}$ yields 30 different terms. Then we have to act with $d^{[4]}=\frac{1}{4!}\eps_{\al\beta\ga\da}d^{\al\beta\ga\da}$, with each derivative capable of hitting each of the six $\phi$'s in each of the 60 terms. Some of the resulting terms vanish as $d^m\phi=0$, but still many remain. Instead of writing them all at once, it helps to organize the terms by their field content. Concentrating on the ``$A$-only'' part of $\mc{H}^{(2)}$, we collect terms with 0 or 4 $d$'s (or $q$'s, which become $d$'s upon projection) remaining on each $\phi$ after $\{d^m,d_n\}=-Z\da^m_n$ is used. We find 
\bea
\mc{H}^{(2)}_{|A^6} &=& \frac{i}{16\sqrt2\cdot 18}\eps_{m n p q}\eps^{m r s t}
\frac{1}{4!}\eps^{i j k l}\frac{1}{4!}\eps_{\al\beta\ga\da}
(-4)\p^{+}\phibar\cdot\frac{1}{\p^{+}}(\p^{+}\phibar\cdot d^\al q^{n p q}\phibar) \nn\\
&\cdot& \frac{1}{\p^{+3}}\Big[ 
4\p^{+}d^{\beta\ga\da}d_{i j k}\phi\cdot\frac{1}{\p^{+}}(\p^{+}\phi\cdot d_l q_{r s t}\phi) \nn\\
&&\qquad +3\times6\p^{+}d^{\beta\ga}d_{i j}\phi\cdot\frac{1}{\p^{+}}
(2\p^{+}d^\da d_k\phi\cdot d_l q_{r s t}\phi
+\p^{+}\phi\cdot d^\da d_{k l}q_{r s t}\phi) \nn\\
&&\qquad +3\times4\p^{+}d^\beta d_i\phi\cdot\frac{1}{\p^{+}}
(3\p^{+}d^{\ga\da}d_{j k}\phi\cdot d_l q_{r s t}\phi
-2\times 3\p^{+}d^\ga d_j\phi\cdot d^\da d_{k l}q_{r s t}\phi \nn\\
&&\hspace{200pt}
+\p^{+}\phi\cdot d^{\ga\da}d_{j k l}q_{r s t}\phi) \nn\\
&&\qquad +\p^{+}\phi\cdot\frac{1}{\p^{+}}(4\p^{+}d^{\beta\ga\da}d_{i j k}\phi\cdot d_l q_{r s t}\phi
+3\times 6\p^{+}d^{\beta\ga}d_{i j}\phi\cdot d^\da d_{k l}q_{r s t}\phi \nn\\
&&\hspace{85pt}
+3\times 4\p^{+}d^\beta d_i\phi\cdot d^{\ga\da}d_{j k l}q_{r s t}\phi
+\p^{+}\phi\cdot d^{\beta\ga\da}d_{i j k l}q_{r s t}\phi)\Big] \ ,
\eea
where we omitted $f^{a b c d}f^{a b^\rr c^\rr d^\rr}$ while keeping the order of $\phi$'s fixed. After a bit of algebra, we find that the 10 terms inside the square bracket combine into the $\p^{+3}$ derivative of a single term. Rewriting the result in terms of $A$'s, we obtain
\bea
\mc{H}^{(2)}_{|A^6}=4f^{a b c d}f^{a b^\rr c^\rr d^\rr}
\Abar_b\cdot\frac{1}{\p^{+}}(\Abar_c\cdot\p^{+}A_d)
\cdot A_{b^\rr}\cdot\frac{1}{\p^{+}}(A_{c^\rr}\cdot\p^{+}\Abar_{d^\rr}) \ .
\eea
Comparing this with the corresponding part in (\ref{HX}), we find that the two expressions agree.~\footnote{
The ``$A$-only'' part of the third line in (\ref{HX}) vanishes identically. Therefore, it would be incorrect to argue that the matching of (\ref{Hqf2}) with (\ref{HX}) for the ``$A$-only'' part, together with the $SO(8)$ $R$-symmetry of $\mc{H}^{(2)}$, guarantees that they match for other parts as well.
}
Turning to the ``$C$-only'' part of (\ref{Hqf2}), we find that
\bea
\mc{H}^{(2)}_{|C^6} &=& \frac{i}{16\sqrt2\cdot 18}\eps_{m n p q}\eps^{m r s t}
\frac{1}{4!}\eps^{i j k l}\frac{1}{4!}\eps_{\al\beta\ga\da}
(-2\times6)\p^{+}d^{\al\beta}\phibar\cdot\frac{1}{\p^{+}}(3\p^{+}d^\ga q^n\phibar\cdot q^{p q}\phibar) \nn\\
&\cdot& \frac{1}{\p^{+3}}\Big[
4\p^{+}d^\da d_{i j k}\phi\cdot\frac{1}{\p^{+}}(3\p^{+}d_l q_r\phi\cdot q_{s t}\phi) \nn\\
&&\hspace{25pt}
+6\p^{+}d_{i j}\phi\cdot\frac{1}{\p^{+}}(3\p^{+}d^\da d_{k l}q_r\phi\cdot q_{s t}\phi
-6\p^{+}d_k q_r\phi\cdot d^\da d_l q_{s t}\phi)\Big] \ .
\eea
The three terms inside the square bracket combine into the $\p^{+}$ derivative of a single term, and rewriting the result in terms of $C$'s, we obtain
\bea
\mc{H}^{(2)}_{|C^6}=\p^{+}C_{i j}\cdot\frac{1}{\p^{+}}(\p^{+}C^{n i}\cdot C_{m n})
\cdot\frac{1}{\p^{+2}}\Big[\p^{+}C^{j k}\cdot\frac{1}{\p^{+}}(\p^{+}C_{k l}\cdot C^{l m})\Big] \ .
\eea
The identity (\ref{sixCid}) allows to rewrite this as
\bea
\mc{H}^{(2)}_{|C^6} &=& -\frac{1}{16}\Big\{
\p^{+}C_3\cdot\frac{1}{\p^{+}}(\p^{+}C_2\cdot C_2)\cdot
\frac{1}{\p^{+2}}\Big[\p^{+}C_3\cdot\frac{1}{\p^{+}}(\p^{+}C_1\cdot C_1)
+\p^{+}C_1\cdot C_1\cdot C_3\Big] \nn\\
&&\hspace{20pt}
+\p^{+}C_3\cdot C_1\cdot C_2\cdot\frac{1}{\p^{+2}}\Big[
\p^{+}C_3\cdot\frac{1}{\p^{+}}(\p^{+}C_1\cdot C_2)+\p^{+}C_1\cdot C_2\cdot C_3\Big] \nn\\
&&\hspace{20pt}
+\p^{+}C_1\cdot C_1\cdot C_2\cdot\frac{1}{\p^{+2}}\Big[
\p^{+}C_2\cdot\frac{1}{\p^{+}}(\p^{+}C_3\cdot C_3)+\p^{+}C_3\cdot C_3\cdot C_2\Big] \Big\} \ .
\eea
Each square bracket can be written as a total $\p^{+}$ derivative of a single term. The sum of the 1st and 3rd lines similarly yields a total $\p^{+}$ derivative on the first triplet of $C$'s. Finally, using the following identity
\bea
\p^{+}C_3\cdot C_1\cdot C_2\cdot\frac{1}{\p^{+}}\Big[C_3\cdot\frac{1}{\p^{+}}(\p^{+}C_1\cdot C_2)\Big]
=-\frac{1}{6}C_1\cdot C_2\cdot C_3\cdot C_1\cdot C_2\cdot C_3 \ ,
\eea
we find that
\bea
\mc{H}^{(2)}_{|C^6}=\frac{1}{16}
C_1\cdot\frac{1}{\p^{+}}(\p^{+}C_2\cdot C_2)\cdot C_1\cdot\frac{1}{\p^{+}}(\p^{+}C_3\cdot C_3)
+\frac{1}{96}C_1\cdot C_2\cdot C_3\cdot C_1\cdot C_2\cdot C_3 \ , \quad
\eea
which matches the corresponding part in (\ref{HX}). We also verified that the $\mc{H}_{|A^2 C^4}^{(2)}$ and $\mc{H}_{|A^4 C^2}^{(2)}$ parts of (\ref{Hqf2}) and (\ref{HX}) match as well, giving
\bea
\mc{H}_{|A^2 C^4}^{(2)} &=& \qter A\cdot\frac{1}{\p^{+}}(C_1\cdot\p^{+}C_1)\cdot\Abar\cdot
\frac{1}{\p^{+}}(C_2\cdot\p^{+}C_2)
+\frac{1}{8} A\cdot C_1\cdot C_2\cdot\Abar\cdot C_1\cdot C_2 \nn\\
&+& \qter C_1\cdot\frac{1}{\p^{+}}(C_2\cdot\p^{+}C_2)\cdot C_1\cdot
\frac{1}{\p^{+}}(A\cdot\p^{+}\Abar+c.c.) \ ,
\nn\\
\mc{H}_{|A^4 C^2}^{(2)} &=& A\cdot\frac{1}{\p^{+}}(A\cdot\p^{+}\Abar)\cdot\Abar\cdot
\frac{1}{\p^{+}}(C_1\cdot\p^{+}C_1)+c.c. \nn\\
&+& C_1\cdot\frac{1}{\p^{+}}(\Abar\cdot\p^{+}A)\cdot C_1\cdot
\frac{1}{\p^{+}}(A\cdot\p^{+}\Abar) \ .
\eea
In the $\mc{H}_{|A^2 C^4}^{(2)}$ case, we had to use the identity (\ref{fourCid}). In all the cases, we used the $[b c d]$ antisymmetry of $f^{a b c d}$ and performed various integrations by parts. However, surprisingly, the Fundamental Identity (\ref{FI}) was never needed in this analysis.



\begin{thebibliography}{99}
\renewcommand{\baselinestretch}{1}\normalsize

\bibitem{bl1}
  J.~Bagger and N.~Lambert,
  ``Modeling multiple M2's,''
  Phys.\ Rev.\  D {\bf 75}, 045020 (2007)
  [arXiv:hep-th/0611108].

\bibitem{gus}
  A.~Gustavsson,
  ``Algebraic structures on parallel M2-branes,''
  Nucl.\ Phys.\  B {\bf 811}, 66 (2009)
  [arXiv:0709.1260 [hep-th]].

\bibitem{bl2}
  J.~Bagger and N.~Lambert,
  ``Gauge Symmetry and Supersymmetry of Multiple M2-Branes,''
  Phys.\ Rev.\  D {\bf 77}, 065008 (2008)
  [arXiv:0711.0955 [hep-th]].


\bibitem{gsp}
  J.~Gomis, A.~J.~Salim and F.~Passerini,
  ``Matrix Theory of Type IIB Plane Wave from Membranes,''
  JHEP {\bf 0808}, 002 (2008)
  [arXiv:0804.2186 [hep-th]].

\bibitem{hll}
  K.~Hosomichi, K.~M.~Lee and S.~Lee,
  ``Mass-Deformed Bagger-Lambert Theory and its BPS Objects,''
  Phys.\ Rev.\  D {\bf 78}, 066015 (2008)
  [arXiv:0804.2519 [hep-th]].


\bibitem{nahm}
  W.~Nahm,
  ``Supersymmetries and their representations,''
  Nucl.\ Phys.\  B {\bf 135}, 149 (1978).


\bibitem{lm}
  H.~Lin and J.~M.~Maldacena,
  ``Fivebranes from gauge theory,''
  Phys.\ Rev.\  D {\bf 74}, 084014 (2006)
  [arXiv:hep-th/0509235].


\bibitem{cm}
  S.~R.~Coleman and J.~Mandula,
  ``All Possible Symmetries Of The S Matrix,''
  Phys.\ Rev.\  {\bf 159}, 1251 (1967).

\bibitem{hls}
  R.~Haag, J.~T.~Lopuszanski and M.~Sohnius,
  ``All Possible Generators Of Supersymmetries Of The S Matrix,''
  Nucl.\ Phys.\  B {\bf 88}, 257 (1975).


\bibitem{ss}
  A.~Salam and J.~A.~Strathdee,
  ``Supergauge Transformations,''
  Nucl.\ Phys.\  B {\bf 76}, 477 (1974).

\bibitem{sg}
  W.~Siegel and S.~J.~Gates,
  ``Superprojectors,''
  Nucl.\ Phys.\  B {\bf 189}, 295 (1981).

\bibitem{bln1}
  L.~Brink, O.~Lindgren and B.~E.~W.~Nilsson,
  ``N=4 Yang-Mills Theory On The Light Cone,''
  Nucl.\ Phys.\  B {\bf 212}, 401 (1983).

\bibitem{man}
  S.~Mandelstam,
  ``Light Cone Superspace And The Ultraviolet Finiteness Of The N=4 Model,''
  Nucl.\ Phys.\  B {\bf 213}, 149 (1983).

\bibitem{bln2}
  L.~Brink, O.~Lindgren and B.~E.~W.~Nilsson,
  ``The Ultraviolet Finiteness Of The N=4 Yang-Mills Theory,''
  Phys.\ Lett.\  B {\bf 123}, 323 (1983).

\bibitem{abkr}
  S.~Ananth, L.~Brink, S.~S.~Kim and P.~Ramond,
  ``Non-linear realization of PSU(2,2$|$4) on the light-cone,''
  Nucl.\ Phys.\  B {\bf 722}, 166 (2005)
  [arXiv:hep-th/0505234].

\bibitem{abhs}
  S.~Ananth, L.~Brink, R.~Heise and H.~G.~Svendsen,
  ``The N=8 Supergravity Hamiltonian as a Quadratic Form,''
  Nucl.\ Phys.\  B {\bf 753}, 195 (2006)
  [arXiv:hep-th/0607019].

\bibitem{db}
  D.~V.~Belyaev,
  ``Dynamical supersymmetry in maximally supersymmetric gauge theories,''
  Nucl.\ Phys.\  B {\bf 832}, 289 (2010)
  [arXiv:0910.5471 [hep-th]].

\bibitem{bbkr}
  D.~Belyaev, L.~Brink, S.~S.~Kim and P.~Ramond,
  ``The BLG Theory in Light-Cone Superspace,''
  JHEP {\bf 1004}, 026 (2010)
  [arXiv:1001.2001 [hep-th]].

\bibitem{bn}
  B.~E.~W.~Nilsson,
  ``Light-cone analysis of ungauged and topologically gauged BLG theories,''
  Class.\ Quant.\ Grav.\  {\bf 26}, 175001 (2009)
  [arXiv:0811.3388 [hep-th]].

\bibitem{thooft}
  G.~'t Hooft,
   ``Computation of the quantum effects due to a four-dimensional
  pseudoparticle,''
  Phys.\ Rev.\  D {\bf 14}, 3432 (1976)
  [Erratum-ibid.\  D {\bf 18}, 2199 (1978)].

\bibitem{bvvn}
  A.~V.~Belitsky, S.~Vandoren and P.~van Nieuwenhuizen,
  ``Yang-Mills and D-instantons,''
  Class.\ Quant.\ Grav.\  {\bf 17}, 3521 (2000)
  [arXiv:hep-th/0004186].

\bibitem{lr}
  N.~Lambert and P.~Richmond,
  ``M2-Branes and Background Fields,''
  JHEP {\bf 0910}, 084 (2009)
  [arXiv:0908.2896 [hep-th]].

\bibitem{gnp}
  U.~Gran, B.~E.~W.~Nilsson and C.~Petersson,
  ``On relating multiple M2 and D2-branes,''
  JHEP {\bf 0810}, 067 (2008)
  [arXiv:0804.1784 [hep-th]].

\bibitem{gio}
  A.~S.~Galperin, E.~A.~Ivanov and V.~I.~Ogievetsky,
  ``Grassmann analyticity and extended supersymmetries,''
  JETP Lett.\  {\bf 33}, 168 (1981)
  [Pisma Zh.\ Eksp.\ Teor.\ Fiz.\  {\bf 33}, 176 (1981)].


\bibitem{abjm}
  O.~Aharony, O.~Bergman, D.~L.~Jafferis and J.~Maldacena,
  ``N=6 superconformal Chern-Simons-matter theories, M2-branes and their
  gravity duals,''
  JHEP {\bf 0810}, 091 (2008)
  [arXiv:0806.1218 [hep-th]].


\bibitem{bl4}
  J.~Bagger and N.~Lambert,
  ``Three-Algebras and N=6 Chern-Simons Gauge Theories,''
  Phys.\ Rev.\  D {\bf 79}, 025002 (2009)
  [arXiv:0807.0163 [hep-th]].



\bibitem{mebius}
J.~E.~Mebius,
``A matrix-based proof of the quaternion representation theorem for four-dimensional rotations,''
arXiv:math/0501249v1.



\end{thebibliography}
\end{document}